\documentclass[12pt]{article}
\input epsf
\usepackage{amssymb}
\usepackage{amsmath}
\makeatletter
\@addtoreset{equation}{section}
\makeatother

\def\be{\begin{equation}}
\def\ee{\end{equation}}

\def\ba{\begin{array}}
\def\ea{\end{array}}

\newcommand{\bea}{\begin{eqnarray}}
\newcommand{\eea}{\end{eqnarray}}

\begin{document}
\hfill{}

\vspace{.5cm}
\vskip 1cm

\vspace{10pt}

\begin{center}
{ \Large {\bf  On $N=8$ attractors}}

\

{\bf  Sergio Ferrara$^{1}$ and Renata Kallosh$^2$
 } \\[7mm]

{\small

$^1$ Physics Department, Theory Unit, CERN, 1211 Geneva 23, Switzerland and\\
\vspace{6pt}  
INFN, Laboratori Nazionali di Frascati, Via Enrico Fermi, 40,00044 Frascati, Italy\\
\vspace{6pt}
{\it Sergio.Ferrara@cern.ch}\\ 
\vspace{6pt}
$^2$ Physics Department, Stanford University, Stanford   CA 94305-4060, USA\\
\vspace{6pt} 
 {\it kallosh@stanford.edu}}

\vspace{10pt}

\vspace{24pt}

\underline{ABSTRACT}

\end{center}

We derive and solve the black hole attractor conditions of $N=8$ supergravity by finding the critical points of the corresponding black hole potential. This is achieved by
 a simple generalization of the symplectic structure of the special geometry  to all extended supergravities with $N>2$.
 There are two classes of solutions for regular black holes, one for 1/8 BPS ones and one for the non-BPS. We discuss the solutions of the moduli at the horizon for BPS attractors using $N=2$ language. An interpretation of some of these results in $N=2$, STU black hole context helps to clarify the general features of the black hole attractors.

\vfill

\newpage

\tableofcontents 

\


\section{Introduction}
Regular $N=8$ BPS black holes in four dimensions have a fascinating feature that the entropy formula is known to be given by the square root of the unique Cartan-Cremmer-Julia quartic invariant \cite{Cartan, Cremmer:1979up} constructed from the fundamental $\mathbf{56}$ representation $(p,q)$ of the $E_{7(7)}$ group. This follows from  U-duality  \cite{Kallosh:1996uy}, so that 
\be
{S_{BPS}\over \pi} = \sqrt {J_4(p,q)}\ ,  \qquad J_4>0
\label{BPS1}\ee
For the $N=2$ BPS  extremal black holes the moduli near the horizon are stabilized due to the attractor mechanism \cite{Ferrara:1995ih}-\cite{FGK}  based on special geometry \cite{deWit:1984pk}-\cite{Ceresole:1995ca}.
For $N>2$ the attractor mechanism was also established starting with \cite{FK2,FGK} and was studied in great details in \cite{Andrianopoli:1996ve}-\cite{Andrianopoli:1997wi}.

The $N=2$ BPS as well as non-BPS extremal black holes under certain condition share the common feature that the moduli near the horizon are stabilized due to the attractor mechanism \cite{FGK,Goldstein:2005hq}. Recently various aspects on the non-BPS black holes were studied, see \cite{Sahoo:2006rp} and references therein.

The purpose of this paper is to establish the black hole attractor conditions of $N=8$ supergravity which would respect the U-duality of N=8 supergravity and allow the possibility for BPS and non-BPS solutions. This will allow us to develop some new  connections between the black holes in string theory/supergravity and quantum information theory (QIT),  to continue the recent trend  due to
\cite{Duff:2006uz}, \cite{Kallosh:2006zs}, \cite{Levay:2006kf}.
It has been argued in  \cite{Kallosh:2006zs} on the basis of the connection between the quartic invariant  of the $E_{7(7)}$ group and the STU black hole entropy \cite{Behrndt:1996hu} that the critical point of $N=8$ black hole potential may have a BPS solution, with the entropy given in eq. (\ref{BPS1}), and a  non-BPS solution with the entropy proportional to the square of the absolute value of the quartic invariant, which is negative for the non-BPS case:
\be
{S_{nonBPS}\over \pi} = \sqrt {- J_4(p,q)}\ ,  \qquad J_4<0
\label{nonBPS1}\ee
The argument in support of the entropy formula (\ref{nonBPS1}) in  \cite{Kallosh:2006zs} was based on a specific example and on the duality symmetry. In this paper we will derive
eqs. (\ref{BPS1}) and (\ref{nonBPS1}) in general case by solving the new BPS/nonBPS attractor equations.

We will also construct a simple generalization of the tools of $N=2$  special geometry to all $N>2$. This will allow us, as an application of the general formalism, to find the attractor equations for any $N$. These equations  are closely related to those in $N=2$. In particular in $N=8$ case we will find a set of simple algebraic equations for N=8 BPS and non-BPS black holes. Equations (\ref{BPS1}), (\ref{nonBPS1}) will be derived  and  the basic features of these black holes will be explained. 
An analysis of the attractor equations for the moduli will be performed by using an $N=2$ language and properties of $N=2$ vector multiplets embedded into $N=8$ supergravity.

\section{Generalization  of $N=2$ special geometry for $N>2$}

Here we  present  simplified version of  flat symplectic bundles which were constructed in \cite{Andrianopoli:1996ve} as  a generalization of N=2 special geometry.
Consider any $N\geq 2$,  d=4 supergravity interacting with some vector multiplets. In $N=2$ case the theory is defined by the elegant special geometry \cite{Ceresole:1995jg},
 \cite{Ceresole:1995ca}, based on  symplectic sections $(f,h)$ \footnote{Here, as well as in refs. \cite{Andrianopoli:1996ve}-\cite{Andrianopoli:1997wi} $(f,h)$ are the complex conjugate of those introduced in \cite{Ceresole:1995jg},
 \cite{Ceresole:1995ca}.}:
\be
(f , h)\equiv (L^\Lambda,\,  \overline {\mathcal D}_{\hat {\bar k}} \bar L^\Lambda \, ; M_\Lambda , \, \overline {\mathcal D}_{\hat {\bar k}} \bar M_\Lambda )
\ee
Here the ``hat'' covariant derivative over the moduli ${\mathcal D}_{\hat { k}} $ means the flat derivative in the moduli space. It is related to the ``curved'' derivative over the moduli as follows: ${\mathcal D}_{\hat { k}}= e_{\hat { k}}^k {\mathcal D}_k$ where the inverse bein $e_{\hat { k}}^k$ is such that the metric of the curved moduli space is $G_{k\bar k} = e_k^{\hat { k}} e_{\bar k}^{\hat {\bar  k}} \delta_{{\hat { k}}{\hat {\bar  k}}}$ and $\delta_{{\hat { k}}{\hat {\bar  k}}}$ is the moduli independent flat metric. Here  ${\mathcal D}_k\equiv  \partial_k+{1\over 2}\partial_k K$ where $K$ is the K\"{a}hler potential. Note that $(f,h)$ are square complex matrices.

Let us first introduce a real symplectic $Sp(2n, \mathbb{R})$ matrix
\be \label{Symplectic}
{\mathcal{S}}= \begin{pmatrix}
  A & B \\
  C & D 
\end{pmatrix} \qquad \mathcal{S}^t \;  \Omega \; \mathcal{S}= \Omega \qquad \Omega = \begin{pmatrix}
 0 & -\mathbb{I} \\
 \mathbb{I} & 0
\end{pmatrix}
 \ee
so that
\be
A^t C- C^t A=0 \qquad B^t D- D^t B=0 \qquad A^t D - C^t B=1
\ee
The sections $(f,h)$ are related to the elements of the real $Sp(2n, \mathbb{R})$ matrix  \footnote{For $N=2$, $n=n_v+1$ where $n_v$ is the number of vector multiplets. For $N>2$ $n$ refers to the total number of vectors in the theory, i. g. for $N=8$, $n=28$.} (\ref{Symplectic}) as follows  \footnote{This is a standard element of a flat symplectic bundle \cite{Strominger:1990pd}, \cite{Ceresole:1995ca}.}

\be \label{S}
\begin{pmatrix}
  A & B \\
  C & D 
\end{pmatrix}
\quad \Rightarrow \quad 
\begin{pmatrix}
  f \\
h
\end{pmatrix}=  {1\over \sqrt 2}
\begin{pmatrix}
 A-iB \\
  C-iD
\end{pmatrix}
\ee
and vice versa

\be \label{inv}{\mathcal{S}}= 
\begin{pmatrix}
  A & B \\
  C & D 
\end{pmatrix}=\sqrt 2
\begin{pmatrix}
  {\rm Re}\, f & -{\rm Im} \,f \\
  {\rm Re} \,h  & -{\rm Im}\, h 
\end{pmatrix}
\ee

For the elements of the section $(f,h)$ this means that 
\be
i(f^\dagger h - h^\dagger f)=1 \ , \qquad f^t h - h^t f=0
\ee
where ${}^\dagger$ means hermitian conjugation and ${}^t$ means  transpose. We  introduce the matrix \cite{Ceresole:1995jg} 
\be
\mathcal{N}= hf^{-1}
\ee
A more detailed structure of indices is useful. The flat, tangent ones we denote $a$ whereas the vector indices are $\Lambda$, so that the section is  $(f^\Lambda_a  ,h_{\Lambda a})$ and 
\be
f^t h \rightarrow f_a^\Lambda h_{\Lambda b}\ ,  \qquad f^\dagger h \rightarrow \bar f^{\Lambda a} h_{\Lambda b} \ , \qquad \mathcal{N}_{\Lambda \Sigma} = h_{\Lambda a} (f^{-1})^a_{\Sigma}
\ee
Introducing 
\be
V_a= \begin{pmatrix}
  f^\Lambda_a  \\
  h_{\Lambda a}
\end{pmatrix}
\ee
we find few simple properties which are the generalizations of the special geometry relations.
\be
\langle \,  V_a, V_b \, \rangle= V_a^t \Omega V_b = -f^t h + h^t f=0
\label{general1}\ee
\be
\langle \,  \bar V^a, V_b \, \rangle= V^{a\dagger}  \Omega V_b = -f^\dagger h + h^\dagger f= i\delta^a_b
\label{general2}\ee

{\it Example of N=2 special geometry}

\be
V_a \Rightarrow (V, \, \overline{\mathcal D}_{\hat {\bar k}} \bar V )
\ee
In N=2 special geometry we have $f^\Lambda_a= (L^\Lambda, \overline{\mathcal D}_{\hat {\bar k}}\bar L^\Lambda)$ and $h_{\Lambda a}= (M_\Lambda, \overline{\mathcal D}_{\hat {\bar k}}\bar M_\Lambda= \mathcal{N}_{\Lambda \Sigma } \overline{\mathcal D}_{\hat {\bar k}}\bar L^\Sigma) $ so that $\mathcal{N}_{\Lambda \Sigma }= h_{\Lambda a} (f^{-1})^a_\Sigma$.

In N=2 special geometry the following relations take place.
\be
\langle \,  V, V_{\hat k} \, \rangle=0 \ , \qquad \langle \,  V_{\hat k} , \overline V_{\hat{ \bar k}}\, \rangle = i \delta_{{\hat k} \, {\hat{ \bar k}} }
\ee
\be
i \langle \,  V, \overline V  \, \rangle=1 \qquad  \langle \,  V, \overline V_{\hat {\bar k}}  \, \rangle=0
\ee
If we now take into account that 
\be
V_a= (
 V , \; 
\overline V_{\hat k}
)=\begin{pmatrix}
  f  \\
  h
\end{pmatrix}
\ee
we may recover all relations of N=2 special geometry from our generalized relations given in eqs. (\ref{general1}), (\ref{general2}). 

Let us also remind here the important matrix of the special geometry, $\mathcal{M}(\mathcal{N})$, \cite{Ceresole:1995ca,FK} which plays a significant role in the black hole potential  ${\cal V}_{BH}=- {1\over 2} Q^t\cdot \mathcal{M}(\mathcal{N})\cdot Q$ of N=2 supergravity\footnote{This invariant was called $I_1$ in \cite{FK}.},  where $Q= (p,q)$ is the symplectic charge vector. 
\be
\mathcal{M}(\mathcal{N})=   \begin{pmatrix}
 {\rm Im} \,  \mathcal{N} + {\rm Re}\mathcal{N} \, {\rm Im} \mathcal{N}^{-1} \, {\rm Re}\mathcal{N}   & -{\rm Re}\,  \mathcal{N} {\rm Im} \,  \mathcal{N}^{-1} \\
-{\rm Im}   \mathcal{N}^{-1}\,  {\rm Re}\mathcal{N}\, & {\rm Im} \,  \mathcal{N}^{-1}
\end{pmatrix}  
\ee

{\it General case, $N\geq 2$}

Introduce the new hermitian matrix, 
$
{1\over 2}(\mathcal{M}+i\Omega)
$
and the covariant symplectic vector $\tilde V_a = (\Omega V)_a$ and note that
$
|\tilde V_a \rangle \langle \tilde {\overline V^a }|
$
is hermitian. This leads to the following relation
\be
{1\over 2}(\mathcal{M}+i\Omega)= - (\Omega  V)_a (\Omega  \bar V^a) 
\label{Result}\ee
where
\be
- |\Omega V_a \rangle \langle \Omega {\overline V^a }|= |\Omega V \rangle \langle {\overline V }\Omega |= \begin{pmatrix}
  -hh^\dagger & hf^\dagger \\
 fh^\dagger & -ff^\dagger
\end{pmatrix} = \begin{pmatrix}
 -h_{\Lambda a} \bar h^a_\Sigma  & h_{\Lambda a} \bar f^{a\Sigma} \\
 f^\Lambda_a \bar h^a_\Sigma  & - f^\Lambda_a \bar f^{a\Sigma}
\end{pmatrix} 
\ee
Note that
\be
{1\over 2}(\mathcal{M}+i\Omega) V_a = i \Omega V_a \qquad (\mathcal{M} V_a= i \Omega V_a)
\ee
By multiplying on $Q$ we find
\be
{1\over 2}(\mathcal{M}+i\Omega) Q= - (\Omega  V)_a \langle Q, \bar V \rangle ^a
\label{identity}\ee
Contracting this with $Q^t$ we get
\be
- {1\over 2} Q^t\cdot \mathcal{M}(\mathcal{N})\cdot Q = \langle Q,  V_a \rangle \langle Q, \bar V^a \rangle 
\ee
and 
\bea
\mathcal{M} Q&=& -2 {\rm Re} \, ((\Omega V)_a \langle Q, \bar V^a \rangle )\nonumber \\
\Omega Q&=& -2 {\rm Im} \, ((\Omega V)_a \langle Q, \bar V^a \rangle)
\label{main}\eea
which is a real and imaginary part of our main new identity (\ref{identity}).
These equations together with the condition ${\cal V}'=0$ have a general validity and can be used for studies of the attractors and  general solutions of stabilization equations  in $N\geq 2$ supergravities \footnote{Note that eqs. (\ref{identity}),(\ref{main}) are the generalizations of the equations derived in the literature in \cite{FK,K,Kallosh:2006bt,Bellucci:2006ew}.}.

{\it N=8 case}

Here we have that the  $Sp(56, \mathbb{R})$ matrix ${\mathcal{S}}= \begin{pmatrix}
  A & B \\
  C & D 
\end{pmatrix} $ is a coset representative of ${E_{7(7)}\over SU(8)}$. It is a 56-dimensional representation of $E_{7(7)}$ which is real and symplectic (i. e. the antisymmetric product $(56\times 56)$ contains a  singlet). The sections are given in terms of the following elements, $f_{AB}^{\Lambda \Sigma}$, $h_{\Lambda \Sigma, AB }$   and their complex conjugates, $\bar f^{\Lambda \Sigma, AB}$, $\bar h_{\Lambda \Sigma}^{ AB}$.

The pair of indices $\Lambda\Sigma$ in  $f_{AB}^{\Lambda \Sigma}= - f_{AB}^{\Sigma \Lambda}$ may be taken to  run over the $\mathbf{28}$ of $SL(8, \mathbb{R})$ and in $\mathbf{28}'$ in $h_{\Lambda \Sigma, AB }$ \footnote{Note that other parametrizations are possible, depending on the particular embedding of the $\mathbf{56}$ of $E_{7(7)}$ into $Sp(56, \mathbb{R})$,  \cite{deWit:1982ig,Hull:1984wa,Andrianopoli:2002mf,Hull:2002cv}.}
The pair of indices $AB$ in $f_{AB}^{\Lambda \Sigma}= - f_{BA}^{\Lambda \Sigma}$, run over the $\mathbf{28}$ of $SU(8)$ both for $f$ and $h$ but in $\overline{ \mathbf{28}}$ for the $\bar f$ and $\bar h$.  By writing 
\be
V_a = \begin{pmatrix}
 f_{AB}^{\Lambda \Sigma}\\
h_{\Lambda \Sigma, AB }
\end{pmatrix} \qquad \rm {we \; see \; that} \quad \Lambda\rightarrow [\Lambda \Sigma] = 1, \dots , 28; 
\quad a\rightarrow[AB], = 1, \dots , 28
\ee
The vector coupling matrix is 
\be
\mathcal{N}_{\Lambda\Sigma, \Gamma \Delta}= h_{\Lambda\Sigma, A B} (f^{-1})^{AB}_{\Gamma \Delta}
\ee
The central charge matrix is
\be
Z_{AB}= f_{AB}^{\Lambda \Sigma}e_{\Lambda \Sigma} - h_{\Lambda \Sigma, AB } m^{\Lambda \Sigma}
\equiv \langle Q, V_{AB} \rangle \ee
where the symplectic charge matrix-vector  $Q$ for N=8 consists of electric $e_{\Lambda \Sigma}$ and magnetic $m^{\Lambda \Sigma}$ charges forming the fundamental representation of $E_{7(7)}$
\be
Q \equiv (m^{\Lambda \Sigma}, e_{\Lambda \Sigma} )
\ee

\section{N=8 black hole potential and its critical points}

N=8,  d=4 supergravity \cite{Cremmer:1979up} has the following black hole potential \cite{Andrianopoli:1996ve,FGK}
\be
{\cal V} _{BH}(\phi, Q) = Z_{AB}  Z^{*AB} = \langle Q,  V_{AB} \rangle \langle Q, \bar V^{AB} \rangle \qquad A,B=1,\dots , 8.
\label{N8pot}\ee
Here $Z_{AB}$ (and its conjugate $ Z^{*AB}$) is the central charge matrix (and its conjugate). 
\be
Z_{AB}(\phi, Q)= \langle Q,  V_{AB}\rangle = f_{AB}^{\Lambda \Sigma}e_{\Lambda \Sigma} - h_{\Lambda \Sigma, AB } m^{\Lambda \Sigma}
\label{CH}\ee
where $Q$ is charge vector, a fundamental $56$ of $E_{7(7)}$ and the bein $f_{AB}^{\Lambda \Sigma}(\phi), h_{\Lambda \Sigma, AB } (\phi)$ is an element of the coset space ${E_{7(7)}\over SU(8)}$ connecting
 the  real $56$,  to complex $28$ of $[AB]$. It depends on 70 real scalars $\phi^i$, where the local $SU(8)$ symmetry was used to remove 63 scalars from the 133-dimensional representation of  scalars in $E_{7(7)}$. Summation over vector indices $AB, \Lambda \Sigma$ is understood for $A<B$ and $\Lambda<\Sigma$ in eqs. (\ref{N8pot}), (\ref{CH}).

The covariant derivative of the central charge  is defined by the Maurer-Cartan equations for the coset space:
\be
{\mathcal D}_i Z_{AB}={1\over 2} P_{i,[ ABCD]}(\phi)  Z^{* CD}(\phi, Q)
\label{derivative}\ee
Here $P_{i,[ ABCD]} d\phi^i$ is the  $70\times 70$  vielbein of the ${E_{7(7)}\over SU(8)}$ coset space, $i=1, \dots , 70$ and it is self-dual real
\be
P_{i,[ ABCD]}= {1\over 4!} \epsilon_{ABCDEFGH}(P_{i,[ ABCD]})^*
\label{self}\ee
Here $D_i$ is the SU(8) covariant derivative \cite{Andrianopoli:1996ve}.
Thus the derivative of the black hole potential over 70 moduli is given by the following expression
\be
\partial_i {\cal V}= {1\over 2}\Big ({\mathcal D}_i Z_{AB}  Z^{*AB} + Z_{AB} {\mathcal D}_i  Z^{*AB}\Big )
\ee
Using eq. (\ref{derivative}) we find
\be
\partial_i {\cal V}={1\over 4} \Big ( P_{i,[ ABCD]} Z^{*AB} Z^{*CD} + P_{i}^{[ ABCD]}Z_{AB} Z_{CD}\Big )
\ee
and with account of the self-duality condition (\ref{self}) we get
\be
\partial_i {\cal V}={1\over 4}  P_{i,[ ABCD]}\Big [Z^{*[CD} Z^{*AB]} + {1\over 4!} \epsilon ^{CDABEFGH} Z_{EF} Z_{GH} \Big ]
\label{critical1}\ee

Now the crucial observation helps to find and extremely simple {\it algebraic} expression for the critical points of all regular black holes by ``flattening its indices''.   The $70\times 70$-bein $P_{i,[ ABCD]}$ is invertible. This means that we can multiply eq. (\ref{critical1}) on $(P_{i,[ A'B'C'D']})^{-1}$ and we get a necessary and sufficient condition for the critical points of the black hole potential with regular $70\times 70$-beins:
\be
Z^{*[AB}Z^{*CD]}+ {1\over 4!} \epsilon ^{ABCDEFGH} Z_{EF} Z_{GH}=0
\label{N8}\ee
To solve these equations we can use the $SU(8)$-symmetry and work in the canonical basis for the antisymmetric central charge matrix \cite{Ferrara:1980ra} where it has only the non-vanishing complex eigenvalues $z_1=Z_{12}, z_2=Z_{34}, z_3= Z_{56}, z_4=Z_{78}$. In this basis the attractor equations  are
\bea
&& z_1 z_2 + z^{*3}z^{*4}=0\nonumber\\
&& z_1 z_3 + z^{*2}z^{*4}=0\nonumber\\
&& z_2 z_3 + z^{*1}z^{*4}=0
\label{attractors}\eea
The $SU(8)$ symmetry allows to bring all 4 complex eigenvalues to the following normal form \cite{Ferrara:1980ra} 
\be
z_i= \rho_i e^{i\varphi/4} \qquad i=1,2,3,4.
\ee
so that only 5 real parameters are  independent\footnote{The black hole solutions with 5 charges have been constructed in \cite{CT2}.}, 4 absolute values $\rho_i$ and an overall phase, $\varphi$ since the relative phase of each eigenvalue can be changed but not the overall phase.
The quartic $J_4$ invariant in this basis acquires the following form \cite{FM}
\be
J_4=\Big [ (\rho_1+ \rho_2)^2- (\rho_3+ \rho_4)^2\Big] \Big [ (\rho_1- \rho_2)^2- (\rho_3- \rho_4)^2\Big]+ 8 \rho_1 \rho_2 \rho_3 \rho_4 (\cos \varphi -1)
\ee
Without loss of generality we may order the four moduli of eigenvalues as follows
\be
\rho_1\geq \rho_2\geq \rho_3\geq \rho_4
\ee
so that the first term in $J_4$ is positive, null or negative depending whether $\rho_1- \rho_2 \geq \rho_3- \rho_4$ or $\rho_1- \rho_2 \leq \rho_3- \rho_4$. The last term is negative or null (it is null if one of the eigenvalues is vanishing or $\varphi=0$).

As we will see below, in terms of the central charge eigenvalues
\be
Z_{AB}= \begin{pmatrix}
  \rho_{1} & 0 & 0 & 0 \\
 0 & \rho_{2} & 0 & 0 \\
  0 & 0 & \rho_{3} & 0 \\
 0& 0 & 0 & \rho_{4} 
\end{pmatrix}\otimes \begin{pmatrix}
  0 & 1 \\
 -1 & 0 
\end{pmatrix} e^{i\varphi/4}
\label{ZAB}\ee
the 1/8 BPS states attractors correspond  to $\rho_2= \rho_3=\rho_4=0$ and the non-BPS one to $\rho_i=\rho$ and $\varphi=\pi$.

Note that for the non-BPS critical points of the potential the matrix of the second derivative may not be positive definite \cite{Goldstein:2005hq}  and a critical point of the potential may not be its minimum, in such case it could be repeller rather than an attractor of the motion.  This analysis will be done elsewhere.

N=8 attractor equations (\ref{attractors}) have 2 solutions for regular black holes 
\begin{enumerate}
  \item 1/8 BPS solution 
  \be
  z_1= \rho_{BPS} e^{i\varphi_1} \neq 0 \qquad z_2=z_3=z_4=0 \qquad J_4^{BPS}=\rho_{BPS}^4>0
  \ee
  The black hole entropy-area of the BPS black holes with 1/8 of N=8 unbroken supersymmetry is given by
\be
{S_{BPS}(Q)\over \pi} = {A_{BPS}(Q)\over 4 \pi}= \sqrt {J_4^{BPS}(Q)}= \rho_{BPS}^2
\label{BPS}\ee
  \item non-BPS
  \be
z_i= \rho \, e^{i{\pi\over 4}}  \qquad J_4^{nonBPS}= -16 \rho_{nonBPS}^4
  \ee
    The black hole entropy-area of the non-BPS black holes all supersymmetries broken is given by
\be
{S_{nonBPS}(Q)\over \pi} = {A_{nonBPS}(Q)\over 4 \pi}= \sqrt {-J_4^{BPS}(Q)}= 4 \rho_{nonBPS}^2
\label{nonBPS}\ee
\end{enumerate}
The deep meaning of the extra factor 4 in the non-BPS solution as compared with BPS one will be clear when we will present the N=2 interpretation of the N=8 result.

Various examples of regular BPS and non-BPS $N=8$ black holes are known in terms of quantized charges. In a series of papers  \cite{Khuri:1995xq}, \cite{Ortin:1996bz}
multiple examples of BPS and non-BPS extremal black holes in various four-dimensional supergravities have been discovered. In \cite{Khuri:1995xq} it was shown that the extremal dyonic Reissner-Nordstr\"{o}m black hole can be a non-BPS solution of N=8 supergravity. In our case this is $z_i= \rho \, e^{i{\pi\over 4}}= q+i p= {1\over \sqrt 2}|q| (1+i)$. The same solution was discovered in \cite{FM} in the context of D0-D6 solution dual to KK-monopole with KK-momentum. This solution has been identified in \cite{Kallosh:2006zs} in the setting of QIT as the canonical Greenberger,  Horne,  Zeilinger
(GHZ) state \cite{Greenberger}, see Fig. 10 in \cite{Kallosh:2006zs}. The contribution to the entropy comes from a single term in $J_4$ which is always negative, it is equal to $- Tr (xy)^2 $ in notation of \cite{Kallosh:2006zs}, and therefore it can only be a non-BPS state.

Another class of   solutions of $N=8$ theory were presented in \cite{Ortin:1996bz} in the context of the no-axion STU model and it was realized there that the flip of the sign of one of the charges relates a 1/8 BPS solution to a non-BPS extremal solution, as was also noticed in more general Calabi-Yau setting in \cite{Goldstein:2005hq,Kallosh:2006bt} and as we have found here  in general  $N=8$ setting. The contribution to the entropy comes from a single  term in $J_4$ of the form
  $-4 {\mbox P\hskip- .1cm f}~ x$,  in notation of \cite{Kallosh:2006zs},  which is a product of 4 charges, e. g. $x^{12}, x^{34}, x^{56}, x^{78}$.  For example,  if $x^{12} x^{34} x^{56} x^{78} < 0$ and $J_4 >0$ we have a BPS combination of charges. If the sign of one of these charges is reversed, we find a non-BPS solution.  In the context of QIT this state was qualified in \cite{Kallosh:2006zs} as the generalized GHZ class of states, see Fig. 9 there.

\section{Generalized special geometry at the attractor point}
The identity (2.20) which generalizes the special geometry formula of N=2 theory can be further simplified at the attractor points. For 1/8 BPS attractors, using eq. (\ref{ZAB}) we get for $Z_{12}=\rho_{BPS}e^{i\varphi_{BPS}}$.
\be
{1\over 2} (\mathcal{M}+i\Omega) Q= - (\Omega V)_{AB}  Z^{*AB} = - (\Omega V)_{12} Z^{*12} 
\label{BPSN8}\ee 
If we pick up a particular component of the $\mathbf{56}$-vector $[(\mathcal{M}+i\Omega) Q]_0$ we get the ratio
\be
{ (\mathcal{M}+i\Omega) Q\over [(\mathcal{M}+i\Omega) Q]_0]} = {(\Omega V)_{12} \over  [(\Omega V)_{12}]_0}
\label{fixed}\ee 
which expresses the scalars in the right hand side  as the function of quantized charges $Q$.

This is in a complete analogy with the N=2 attractors. It was shown in eqs. (42), (44) of \cite{FK} that special geometry requires at $DZ=0$ that
\be
2i \bar Z L^\Sigma = p^\Sigma +i {\partial I_1(p, q)\over \partial q_\Sigma} \ , \qquad 2i \bar Z M_\Sigma = q_\Sigma -i {\partial I_1(p, q)\over \partial p^\Sigma}
\label{FK1}\ee
The symplectic invariant $I_1$  in the arbitrary point of the moduli space is given by the following formula: $I_1(z, \bar z; p,q)\equiv |Z|^2 + |DZ|^2$.
In eq. (\ref{FK1}) $I_1(p, q)$ is taken
at the BPS attractor point $DZ=0$ where $z=z(p,q)$ and $\bar z=\bar z(p,q)$ so that
\be
I_1(p, q)\equiv  I_1(z, \bar z; p,q)|_{DZ=0} \ .
\ee 
This gives the stabilization eqs. in the form
\be
t^\Lambda=
{X^\Lambda\over X^0}=
{p^\Lambda+i {\partial I_1(p, q)\over \partial q_\Lambda}\over p^0+i {\partial I_1(p, q)\over \partial q_0}} \ , 
\qquad 
t_\Lambda
=
{F_\Lambda\over F_0}
=
{q_\Lambda-i {\partial I_1(p, q)\over \partial p^\Lambda}\over q_0-i {\partial I_1(p, q)\over \partial p^0}}
\label{stab}\ee
These are the attractor values of special coordinates $t^\Lambda(p,q)$ and their dual, $t_\Lambda(p,q)$, as explicit function of charges for all cases when the entropy formula $ S(p,q)= \pi I_1(p,q)$ is known.
This form of stabilization equations was used by  Bates and Denef in \cite{Bates:2003vx}  for the N=2 multicenter black hole constituent solutions in cases that the black hole entropy is known as the function of charges. One has to replace the charges by harmonic functions to find the multi-center solutions.
Here we see that eqs. (\ref{BPSN8}), (\ref{fixed}) are the N=8 analogs of N=2 eqs.  (\ref{FK1}),  (\ref{stab}).

There is actually a simple way to recast the equation (\ref{fixed})in the language of equations (\ref{FK1}),  (\ref{stab}). Before doing this let's first recall that, following the analysis of \cite{Andrianopoli:1997pn,Andrianopoli:1997wi} for the 1/8 BPS, $N=8$ attractors 30 scalars get fixed and 40 are not fixed. This corresponds to the decomposition of $N=8$ into $N=2$ multiplets, resulting from the decomposition $SU(8)\rightarrow SU(2) \times SU(6)\times U(1)$ of the R-symmetry, \cite{Andrianopoli:1997wi}. Under this splitting one obtains that 70 scalars decompose into 15 complex scalars of 15 vector multiplets  and 40 real scalars of 20 (half)-hypermultiplets.
\be
\mathbf{70}\Rightarrow (\mathbf{15}, \mathbf{1})+ ( \overline {\mathbf {15}}, \mathbf{1})+ (\mathbf{20}, \mathbf{2})
\ee
under $SU(6)\times SU(2)$.
$N=2$ ensures that 40 scalars are not fixed and 30 are fixed, as proved in \cite{Andrianopoli:1997pn}. The 15 complex scalars belong to the submanifold ${SO^*(12)\over U(6)}$ of ${E_{7(7)}\over SU(8)}$. Indeed, the vector and hypermultiplet splitting correspond to two different decompositions of $E_{7(7)}$ with respect to maximal subgroups \cite{Gilmore},
\bea
E_{7(7)} &\rightarrow& SO^*(12) \times SU(2)\\
E_{7(7)} &\rightarrow& E_{6(2)} \times U(1)  
\eea 
where $15_c$ parametrize ${SO^*(12)\over U(6)}$ and $40$ real scalars parametrize ${E_{6(2)}\over SU(2)\times SU(6)}$. Note that the two factors $SU(2), U(1)$ in the two different $E_{7(7)}$ decompositions simply mean that vector multiplet scalars are $SU(2)$ singlets while the hypermultiplet scalars are $U(1)$ singlets. Of course the two subgroups $SO^*(12), E_{6(2)}$ do not commute, otherwise ${E_{7(7)}\over SU(8)}$ would be a product space. Nevertheless, if we disregard the hyperscalars then
\be
{E_{7(7)}\over SU(8)} \Rightarrow {SO^*(12)\over U(6)}
\ee 
which is a symmetric special manifold \cite{Cremmer:1984hc}. Special coordinates for this manifold are obtained by its lifting to 5 dimensions to ${SU^*(6)\over Usp(6)}$ which is a 14-dimensional real (very special) manifold \cite{Ferrara:1997uz,Gunaydin:1983rk}.

The $d=4$ prepotential of the effective $N=2$ supergravity\footnote{We will present a more detailed discussion of this model in future work.}
\be
F(X)={1\over 2^3\cdot 3!} \epsilon_{\Lambda \Sigma \Gamma \Delta \Pi \Omega}{X^{\Lambda\Sigma} X^{\Gamma\Delta} X^{\Pi\Omega}\over X^0}= (X^0)^2 f(t^{\Lambda\Sigma}) \qquad \Lambda, \Sigma = 1, \dots, 6.
\label{prep}\ee
where $t^{\Lambda\Sigma}= -t^{\Sigma\Lambda}= {X^{\Lambda\Sigma}\over X^0}$ are complex coordinates in the $\mathbf{15}$ of $SU^*(6)$ $X^0$ is the graviphoton, corresponding to the decomposition of $SO^*(12)\rightarrow SU^*(6)\times SO(1,1)$ under which the $SO^*(12)$ spinorial $\mathbf{32}$ representation which represent the N=2 electric and magnetic charges (including graviphoton) decomposes as 
\be
\mathbf{32}\rightarrow \mathbf{15}+ \mathbf{15}' + \mathbf{1} + \mathbf{1}'
\ee
Note that in this formulation only 32 of the 56 charges are retained and the symplectic sections $f^{\Lambda\Sigma}, h_{\Lambda\Sigma}$ are identified with the sections  $e^{K/2} (X^{\Lambda\Sigma}, \overline {\mathcal D}_{\hat {\bar k}} \overline X^{\Lambda\Sigma} \, ; F_{\Lambda\Sigma} , \, \overline {\mathcal D}_{\hat {\bar k}} \overline F_{\Lambda\Sigma} )$
where $F_{\Lambda\Sigma}= {\partial F\over \partial X^{\Lambda\Sigma} }$ and special coordinates are adopted.

Here  the  $\mathbf{56}$ of $E_{7(7)}$  is decomposed under $SO^*(12)\times SU(2)$ into $ (\mathbf{32}, \mathbf{1}) + (\mathbf{12}, \mathbf{2})$. 
This is a different embedding of $\mathbf{56}$ into $Sp(56, \mathbb{R})$  compared with \cite{Cremmer:1979up,Hull:1984wa}.  There $\mathbf{56}$ was decomposed into  $ \mathbf{28} + \mathbf{28}'$ of $SL (8, \mathbb{R})$, as we discussed in sec. 2. Here instead the electric-magnetic  splitting corresponds to a further decomposition of $SO^*(12)$ into $ SU^*(6)\times SO(1,1)$, leading to the prepotential in eq. (\ref{prep}). The 32 electric and magnetic charges correspond to the 15 vector multiplets and the graviphoton. The other {\it 24 remaining charges correspond to the vectors of the  6 gravitino multiplets  each of them containing  2 vectors}: $6({3\over 2}, 2(1), {1\over 2})$. These are the charges which do not appear in the N=2 setting as they are partners of the 6 gravitino which we truncate.
These charges can be generated by applying to the 32 charges an $SU(8)$ rotation which is not in $SU(6)\times SU(2)\times U(1)$. Indeed
$$\rm dim \; {SU(8)\over SU(6)\times SU(2)\times U(1)}=24$$
which is precisely the missing charges. 
To see that this counting is correct we observe that the N=2,  BPS black hole has a 32 component charge vector in the coset \cite{Ferrara:1997uz} ${SO^*(12)\over SU(6)}$ which has the signature $(30^-, 1^+)$. The 1/8,  56 component,  $N=8$ BPS charge vector is in the coset ${E_{7(7)}\over E_{6(2)}}$ with the signature $(30^-, 25^+)$, \cite{Ferrara:1997uz} \footnote{The coset represents this charge vector for the fixed value of $I_4$ \cite{Andrianopoli:1997wi}.}. We therefore realize that the missing 24 charges are precisely the extra compact directions which correspond to $SU(8)$ rotations. Note that the same reasoning does not apply to the non-BPS black holes since in this case the signature of the coset, ${E_{7(7)}\over E_{6(6)}}$ is $(28^-, 27^+)$.

Having identified the effective N=2 supergravity with the prepotential in eq. (\ref{prep}) describing the $N=8$ 1/8 BPS attractors we would need to find the expression for the entropy as the function of quantized charges so that equations (\ref{fixed}) can be used. One finds that  the entropy of the ${SO^*(12)\over U(6)}$ model is the same as in $N=6$ supergravity where, with a different identification of the central charges we have   \cite{Andrianopoli:1996ve,Andrianopoli:1997pn}
$
Z_{AB} \Rightarrow N=2$ matter charges and  $Z\rightarrow N=2$ central charge, $A, B =1,\dots , 6$. $P_{i, ABCD}\rightarrow $ is a vielbein of ${SO^*(12)\over U(6)}$. In terms of $Z_{AB}, Z$ the unique $SO^*(12)$ quartic invariant is 
\be
{S\over \pi}= {1\over 2} \sqrt{ 4I_2- I_1+ 32 I_3 +4 I_4 + 4I_5}
\ee
where $A_A{}^C\equiv  Z_{AB}\overline Z^{BC}$ and $I_1= (\rm Tr A)^2$, $I_2= \rm Tr(A)^2 $, $I_3= \rm   Re (\rm Pf Z_{AB} Z)$, $I_4= \rm Tr A Z\bar Z$, $I_5 = Z\bar Z Z\bar Z$.
Note that at $N=2$ attractor point $Z_{AB}=0$ and 
\be
{S\over \pi}= \sqrt {I_5}= Z\bar Z
\ee
as expected. The black hole potential in this model is
\be
{\cal V}= \sum_{A<B} Z_{AB} \overline Z^{AB} + Z\bar Z
\ee
The Maurer-Cartan equations \cite{Andrianopoli:1997wi} are identical to the special geometry relations which imply that $Z_{AB}=0$, $|Z|^2= \sqrt{I_4}$ is indeed a BPS attractor point. The $I_4$ invariant is thus a function of the quantized charges $Q_\alpha$ ($\alpha = 1, \dots , 32)$ which form a left spinor of  $SO^*(12)$. Since $I_4$ is moduli independent, 
\be
I_4(Q_\alpha)= \Gamma^{\alpha\beta\gamma\delta} Q_\alpha Q_\beta Q_\gamma Q_\delta
\ee
where $\Gamma^{\alpha\beta\gamma\delta}$ is a numerical tensor, constructed with the  $SO^*(12)$ $\gamma$-matrices which is totally symmetric in the spinor indices. This accomplishes the solution for 1/8 BPS attractors.

{\it Non-BPS case}
 
In N=8 theory we can actually find the non-BPS equations from eq. (\ref{ZAB}) with $z_i^*= \rho_{nonBPS} e^{-i\pi/4}$
\be
{1\over 2} (\mathcal{M}+i\Omega) Q= - (\Omega V)_{\rm TRACE} \;  \rho_{nonBPS} e^{-i\pi/4}  
\label{nBPSN8}\ee 
where 
\be
(\Omega V)_{\rm TRACE}= (\Omega V)_{12} +(\Omega V)_{34}+ (\Omega V)_{56}+ (\Omega V)_{78}
\ee
Again by dividing on some component of the $\mathbf{56}$-vector we get
\be
{ (\mathcal{M}+i\Omega) Q\over [(\mathcal{M}+i\Omega) Q]_0]} = {(\Omega V)_{\rm TRACE} \over  [(\Omega V)_{\rm TRACE}]_0}
\label{fixedNon}\ee 
Thus equations (\ref{fixed}) and  (\ref{fixedNon}) are defining the scalars of BPS and non-BPS N=8 theory at the black hole horizon. The details of the non-BPS solutions will be considered elsewhere.

\section{STU N=2 interpretation of N=8 attractors}
Consider the STU black holes in N=2 supergravity with the prepotential $F=STU$. The corresponding coordinates represent the coset space $\Big ({SU(1,1)\over U(1)}\Big )^3$.

The algebraic  attractor equations of N=8 theory in eq. (\ref{attractors}) can be identified with the corresponding N=2 attractor equations $\partial V_{BH} =0$ \cite{FGK} in the form
\be
2(\mathcal{D}_i Z) \bar Z + i C_{ijk}G^{j\bar m} G^{k\bar k} \bar {\mathcal{D}}_{\bar m} \bar Z  \bar {\mathcal{D}}_{\bar k} \bar Z =0
\ee
under the correspondence:
\bea
z_1&=& i Z\nonumber \\
z_2&=& \overline{\mathcal{D}_{\hat S} Z}\nonumber \\
z_3&=& \overline{\mathcal{D}_{\hat T} Z}\nonumber \\
z_4&=& \overline{\mathcal{D}_{\hat U} Z}
\eea
We recover the extremization of the potential in N=8 theory via the special geometry in N=2.
The BPS case is 
\be
\mathcal{D}_{\hat S} Z= \mathcal{D}_{\hat S} Z=D_{\hat S} Z=0 \qquad Z\neq 0
\ee
which is equivalent to
\be
z_2= z_3= z_4=0 \qquad z_1\neq 0
\ee
The entropy-area at the BPS attractor points is
\be
{S\over \pi}= {A\over 4\pi} = |Z|^2_{BPS}
\ee
The non-BPS case (for $Z\neq 0$)  requires that
\be
|Z|= |D_{\hat S} Z| = |\mathcal{D}_{\hat T} Z|= |\mathcal{D}_{\hat U} Z|
\ee
and 
the entropy-area at the non-BPS attractor point is \footnote{The multiplicative ``renormalization'' between the entropy and the square of the central charge for non-BPS attractors was found in \cite{Bellucci:2006ew}.   The  factor of 4  difference in STU case was found in the BPS and non-BPS free energies in terms of the
image part of the prepotential in \cite{Parvizi:2006uz}.}
\be
{S\over \pi}= {A\over 4\pi} =|Z|^2+ \sum_{i=1}^3| \mathcal{D}_i Z|^2= 4|Z|^2_{nonBPS}
\ee
This explains the origin of the factor 4 in the BPS-non-BPS entropy relation of the N=8 attractor (see eqs. (\ref{BPS}) and (\ref{nonBPS}). 

\

In conclusion, we have derived the attractor equation by means of an identity (\ref{identity}), for generic $N>2$ supergravity in a form resembling the $N=2$ attractor equations. This identity (\ref{identity})  has to be studied together with the condition for the critical point of the black hole potential,  ${\cal V}'=0$ as in N=2 case of new attractors \cite{K},\cite{Kallosh:2006bt},\cite{Bellucci:2006ew}. In N=8 supergravity the critical point of the black hole potential ${\cal V}'=0$ is given by an amazing set of algebraic attractor equations (\ref{N8}), (\ref{attractors}) which clearly produce 2 solutions for regular black holes: one 1/8 BPS and one non-BPS. The scalars are defined via charges in eqs. (\ref{fixed}) and (\ref{fixedNon}). Finally, the attractor equations for the fixed scalars at the BPS attractor points can be studied using the prepotential (\ref{prep}) for the 15 $N=2$ vector multiplets embedded into $N=8$ supergravity.

We have also pointed out some connections between $N=8$ black holes and quantum information theory in the spirit of the recent proposals in \cite{Duff:2006uz}, \cite{Kallosh:2006zs}, \cite{Levay:2006kf}. A particular class of  regular non-BPS black holes in $N=8$  belongs to   a canonical GHZ state, some other black holes  belong to the generalized GHZ class of states, both in BPS and non-BPS case. It is quite impressive that the extremization of the black hole mass with respect to moduli is related to the process of finding the optimal local distillation protocol of a GHZ state from an arbitrary three-qubit pure state, according to \cite{Levay:2006kf}.

\

{\Large {\bf Acknowledgments}}

It is a pleasure to thank  R. D'Auria, J. Bena, E. Gimon, P. Kraus, F. Larsen, P. Levay,  T. Ort\'{\i}n,  B. Pioline and M. Trigiante
for useful conversations.  We are grateful to  participants of 2006  Frascati Winter school on Attractor mechanism for the interest to this work.  The work of
S.F.~has been supported in part by the European Community Human Potential
Program under contract MRTN-CT-2004-005104 ``Constituents, fundamental
forces and symmetries of the universe'', in association with INFN Frascati
National Laboratories and by D.O.E.~grant DE-FG03-91ER40662, Task C. The work of R.K. was supported by
NSF grant PHY-0244728.

\end{document}